\documentclass[aps,twocolumn]{revtex4}
\usepackage{graphicx}
\usepackage{amsmath}
\usepackage{hyperref}
\usepackage{SIunits}
\def\percent{\%} 
 
\makeatletter 
\def\@dotsep{4.5}
\makeatother

\begin{document}
\title{A toy model of polymer stretching}

\author{Carlo Guardiani}
\email{carlo.guardiani@unifi.it}
\affiliation{Centro Interdipartimentale per lo Studio
 delle Dinamiche Complesse (CSDC), Universit\`a di Firenze, Via Sansone 1, I-50019, Sesto Fiorentino, Italy}
\affiliation{Istituto Nazionale di Fisica Nucleare (INFN), sez.\ Firenze  } 
\author{Franco Bagnoli}
\email{franco.bagnoli@unifi.it}
\affiliation{Dipartimento di Energetica, Universit\`a di Firenze,Via S.~Marta 3, I-50139 Firenze, Italy}
\affiliation{CSDC and INFN, sez.\ Firenze}

\date{\today}

\begin{abstract}
We present an extremely simplified model of multiple-domains  polymer
stretching in an atomic force microscopy experiment. We portray each
module as a binary set of contacts and decompose the system energy 
into a harmonic term (the cantilever) and long-range interactions
terms inside each domain. Exact equilibrium computations and  Monte
Carlo simulations qualitatively reproduce the experimental saw-tooth
pattern of force-extension profiles, corresponding (in our model) to
first-order phase transitions.  We study the influence of the coupling
induced by the cantilever and the pulling speed on the relative
heights of the force peaks. The results suggest that the increasing
height of the critical force for subsequent unfolding events is an
out-of-equilibrium effect due to a finite pulling speed. The
dependence of the average unfolding force on the pulling speed is
shown to  reproduce the experimental logarithmic law. 
\end{abstract}

\maketitle

\newpage

\section{Introduction}

The last decade has witnessed a significant advancement of single molecule
manipulation and visualization techniques~\cite{Review-1,Review-2,Review-3,
Review-4,Review-5,Review-6,Review-7} providing access to the distribution 
of physical properties across many individual molecules and not just average 
properties as was the case of traditional biochemical techniques.

Optical Tweezers (OT) and Atomic Force Microscopy (AFM) in particular, enable 
the study of mechanical properties of proteins such as those in the 
extracellular matrix, in the cytoskeleton and in the muscle, that  
\emph{in vivo} are exposed to stretching forces. The mechanical
properties of macromolecules obtained in these experiments may be
directly connected to corresponding thermodynamical
quantities~\cite{bustamante} with a bit of caution, since mechanical
experiments are usually out-of-equilibrium, as discussed in this
paper. On the other hand, the mechanical properties of biomolecules
may be of direct importance for what concerns bendability (DNA),
translocation of nucleic acids and proteins across cellular membranes,
rigidity and elasticity (structural proteins).


In a force-measuring AFM experiment, the tip of a microscopic cantilever is 
pressed against a flat, gold-covered substrate, coated with a thin layer of 
protein. Upon retraction of the substrate, a protein molecule that may have 
been adsorbed to the cantilever tip is then stretched. Finally, the force the
protein opposes to the stretching is computed from the cantilever deflection, 
and a force-extension plot is drawn. In a typical stretching experiment the 
force-extension curve shows a saw-tooth pattern, each peak corresponding to 
the unfolding of a single domain. However, if the protein is composed by several
different modules, it is difficult to associate each peak to the corresponding
domain. Moreover, in order to neglect the tip-substrate interaction, it is
necessary to have a sufficiently long protein. A common strategy proposed
in the literature to address these problems is to use an engineered protein 
composed by several tandem repeats of the same kind of domain. 
A typical choice is to use domains belonging to the 
immunoglobulin~\cite{Ig}, fibronectin III~\cite{FN-III} or 
cadherin~\cite{cadherin} superfamilies, characterized by 
$\beta$-sandwich structures. Alpha-helical domains, such as those of the 
cytoskeletal protein spectrin~\cite{spectrin}, have also been used. 

Many data about the mechanical properties of modular proteins can be
extracted from the force-extension profiles. The amplitude of each
peak, in  fact, represents the mechanical stability of the
corresponding domain, while  the spacing between peaks reflects the
length of the unfolded domain and is thus proportional to the number
of amino-acids it comprises. The experimental data  also show a
logarithmic dependence of the height of the peaks on the pulling
velocity, so that higher forces are required for domain unraveling
when the pulling occurs very quickly.

\begin{figure}[t]
\begin{center}
\includegraphics[height=\columnwidth,angle=-90]{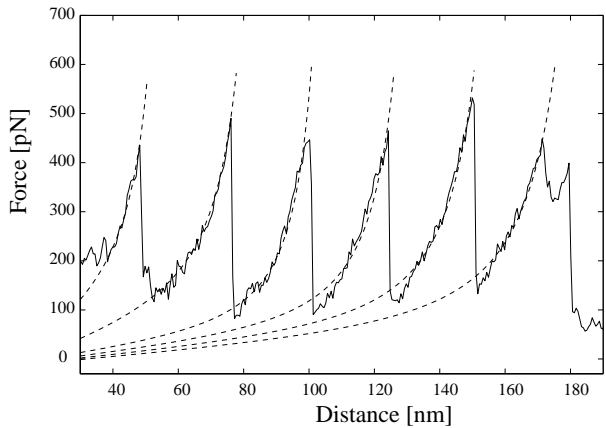}
\end{center}
\caption[The force-extension profile of AFM experiments.]{
The force-extension profile produced in  a typical AFM stretching experiment 
performed on a modular polyprotein  composed by 8 tandem repetitions  of a 
Ig-like domain from titin. The construct is terminated by 2 cystein residues  
expressly introduced to form covalent bonds with a gold surface.  The experiment 
was performed at a constant speed of $200\nano\metre\per\second$. 
Notice that only 7 unfolding peaks 
appear in the plot because the cantilever tip, by chance, established a contact 
with the second domain of the molecule, skipping the first one.
}
\label{figVassalli}
\end{figure}

The features of the force-extension curves (Figure
~\ref{figVassalli})  are accounted for by the entropic elasticity of
polymer chains in solution. The entropy of a polymer, in fact,  is
maximal in the random coil state, whereas it tends to zero in the
fully extended  conformation. The force required to stretch a polymer
thus reflects the entropy  loss and it grows in a non-linear way as
the molecule is lengthened.  The entropic elasticity of biopolymers is
currently modeled through the worm-like  chain  (WLC)~\cite{WLC} and
freely-jointed chain (FJC)~\cite{FJC} models.  The WLC model describes
a molecular chain as a deformable rod whose stiffness is  determined
by the persistence length $p$ (the length scale over which the polymer
loses orientation order). The functional relation between the 
external force and the fractional extension $z/L$ ($z$ is the
end-to-end  distance and $L$ is the contour length) is approximatly
given by the interpolating formula
\begin{equation}
\label{WLC}
 F = \frac{k_{B}T}{p} \left ( \frac{1}{4(1 - z/L)^{2}} - \frac{1}{4} + 
\frac{z}{L} \right ),
\end{equation}
where $k_{B}$ is the Boltzmann constant and $T$ is the temperature.

In the FJC model, the polymer is portrayed as a chain of rigid
segments linked  by frictionless joints so that each segment can point
in any direction irrespective of the orientations of the others. A
measure of the stiffness of  the chain is represented by the Kuhn
segment length $b$ (the average length of the segments). The analytic
relation between the average end-to-end distance $\langle z \rangle$
and the stretching force $F$ is
\begin{equation}
\label{FJC}
 \langle z \rangle = L \left ( \coth \frac{Fb}{k_{B}T} - \frac{k_{B}T}{Fb} 
\right ).
\end{equation}

In our model, the polymer  is described by an array of binary
variables representing native contacts that can be in either of two
states: formed or broken.  The cantilever, on the other hand, is
modeled as a harmonic spring in series with the molecule. The energy
of  the system is the sum of a harmonic term and a term of long-range
interaction modeling in a bulk, coarse-grained way, the chemical
interactions stabilizing  the native conformation of the protein. In
fact, rather than providing a  detailed description, we account for
chemical interactions through a folding  prize attributed to a domain
when the fraction of intact contacts is above  a threshold. We assume
any contact breakdown to produce an equal increment in  the molecule
length.

Let us consider first the force-extension characteristic of our model
in the absence of folding prize.  Let $F$ be a constant pulling force,
$n$ the number of broken contacts and $a$ the length increment per
cleaved contact. If no folding prize is attributed to the molecule in
a native-like state, the Hamiltonian of the system is
$\mathcal{H} = -Fna$ and the partition function is
\[
 Z = \sum_{n} \left ( \begin{array}{c}
     N \\
     n
     \end{array} \right )  e^{\beta Fna} = (1 + e^{\beta Fa})^N. 
\]
Notice that this is also the partition function of an Ising-like
model  in one dimension without coupling among spins. The average
end-to-end distance can thus be computed as 
\begin{equation}
\label{ISING}
\langle z \rangle = a  \langle n \rangle = 
\frac{k_BT}{Z}\frac{\partial Z}{\partial F} =
\frac{Na}{2} \left ( 1 + \tanh \left ( \frac{Fa}{2k_{B}T} \right ) \right ) .
\end{equation}

\begin{figure}[t]
\begin{center}
\includegraphics[width=\columnwidth]{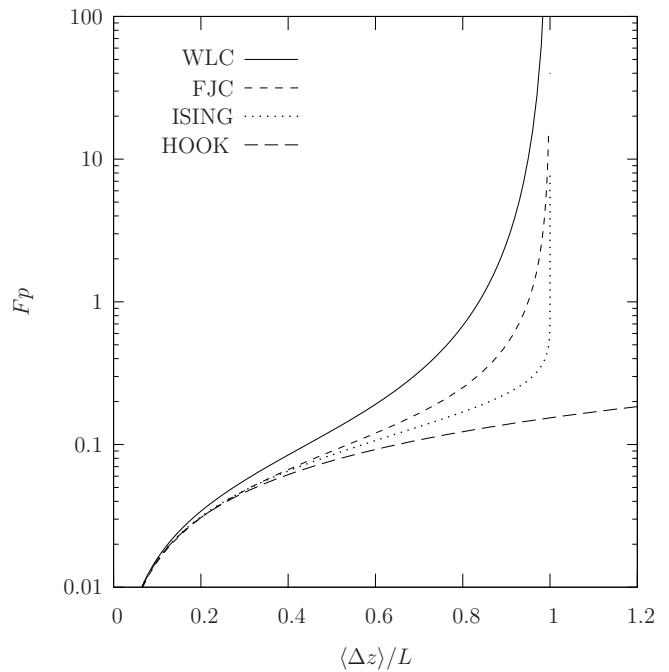}
\end{center}
\caption[Force versus extension for various 
models.]{Force versus extension for various 
models: WLC (interpolated formula) 
with $p=0.4~\nano\metre$, $L=29~\nano\metre$, $T=310~\kelvin$; 
FJC with $b=2p$, ISING with $a=1.3p$ $N=L$, HOOK (harmonic potential)
with $K=1.3p$.
}
\label{models}
\end{figure}

The variation of length $\langle \Delta z \rangle$ versus $F$ in the
WLC,  FJC and our (ISING) models is shown in Figure~\ref{models}, where
we have adjusted the constant $a$ (corresponding to persistence length
$p$ so to make the curve coincide for small elongations.  
Despite the extreme simplicity of our
assumptions, the three curves are qualitatively similar.
This similarities could be increased by adding contact-contact interactions
or considering different elongations for the various contact breaking
events. 

Although the WLC and FJC models more accurately represent the physics
of a polymer, their statistical mechanics treatment is so
complex that one generally employes the average formulas \eqref{WLC} and
\eqref{FJC} for each domain, complemented with an independence
hypothesis of domains and an ad-hoc treatment of the unfolding
event~\cite{Evans-1,Evans-2,Rief}, based on an extension of  Bell's
expression~\cite{Bell} or full Kramer's theory~\cite{Schlierf} 
for the rupture rate coefficient in the 
presence of a time-dependent external force.

\begin{figure}[t]
\begin{center}
\includegraphics[width=\columnwidth]{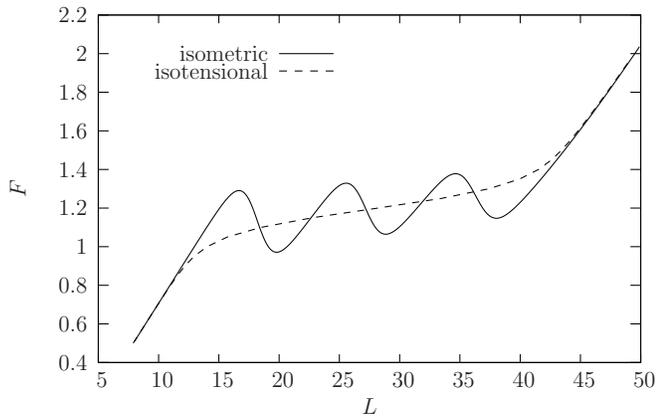}
\end{center}
\caption[Isometric vs isotensional elongations.]{Isometric versus
isotensional elongations for $M=3$, $A=1$, $K=0.1$, $N = 10$, $\theta
= 0.2$ and  $beta = 2$ (see Section~\ref{model} for the illustration
of the model).
}
\label{iso} 
\end{figure}

The
independence assumption is questionable, since all domains are coupled
by the presence of the cantilever. This difference may be explicited
by comparying computations~\cite{bleha} in which the position of the cantilever is
observed (isometric) while the force may fluctuate, 
with computations in which the force is maintained constant
(isotensional). We can obtain exact comparisons of the two different
set-up for our model (see Section~\ref{model}), as shown in
Figure~\ref{iso}. It can be noticed that the isotensional model does
not show any peaks (the peaks are here smoothed due to the small
length of the single module).


Moreover, the peaks in the force-extension profile
(Fig.~\ref{figVassalli}) are a signature of a first-order transition.
As we shall show in the following, this transition naturally arise in
our model due to the long-range coupling (the folding
prize). 

%

Despite its extreme simplicity, our approach captures some important
aspects of the physics of Ig domain stretching. Steered molecular
dynamics simulations performed  in Schulten's group~\cite{Schulten1,
Schulten2, Schulten3}, in fact, showed that the  mechanical unfolding
of the I27 module occurs only after the breakdown of a patch of six
hydrogen bonds bridging the $A'$ and $G$ $\beta$-strands. The rupture
of these critical bonds was shown to be the key event allowing the
full unraveling  of the molecule under an external force. 

Even if this pattern was originally observed in a specific protein, 
it could be hypothesized a more widespread distribution. Makarov and
coworkers~\cite{Makarov}  performed Monte Carlo simulations
of titin forced unfolding. During these simulations the number of
hydrogen bonds at time $t$, $n(t)$, undergoes a random walk.  
It was concluded that a critical value $n^{\#}$ of the
number of hydrogen bond  does exist, such that when $n(t) < n^{\#}$,
the domain unfolds very rapidly.  Makarov also showed that the
number of bonds is roughly one,  when the force is low enough,
whereas for very large pulling rates (and thus large pulling forces),
it is likely to be equal to six, recovering the findings by
Lu and Schulten~\cite{Schulten3}.

The accuracy of the model developed by Makarov and coworkers, allows
quantitative comparisons with experimental data at the expense of very
long simulation times and the need to assume the knowledge of the free
energy profile. Their model is  also based on the hypothesis of
independence of domains which, however, might be incompatible with the
coupling introduced by the cantilever. Since the  transitions shown in
AFM experiments are out of equilibrium~\cite{Rief, Grubmuller},
thermal fluctuations may play a fundamental role. Our model,
conversely, is so  simplified that we can compute exactly the free
energy of a multi-domain protein for the equilibrium case, and perform
long simulations in the out of equilibrium  case.

The model we propose shows interesting similarities to a G\={o} model
with force rescaling. 
%
Recent theoretical studies~\cite{Chan,Cecconi} show that the ability
of G\={o} models to simulate the cooperativity of the folding process
can be enhanced by  imparting an extra energetic stabilization to the
native state so as to  simulate specific interactions appearing only
after the assembly of native-like structures. A rigorous approach to
simulate the stabilizing interactions peculiar of the native state,
would be the use of two different analytic expressions of the
force-field inside and outside the native basin. The same purpose can
be pursued through a much simpler strategy~\cite{Chan}, by rescaling
the conformational force when the fraction of native contacts crosses
a pre-chosen threshold. 

This approach appears to be similar to the one employed in our model.
The energy function of our model comprises an elastic term
and a contact term. The harmonic term accounts for the elasticity of
the cantilever, while the protein can be thought of as a soft
spring so that its contribution to the elasticity of the system is
negligible.

The contact term of our energy function, on the other hand, is a 
stabilizing contribution that the protein receives only when the
fraction of intact contacts crosses a threshold. This approach is thus
equivalent to a force  rescaling occurring whenever the polymer enters
the native basin, and the contact term of the energy function appears
to be a square well.

Finally, from a purely formal point of view, our description of the
polymer  as an array of binary variables, bears some similarity with
the model proposed  by Galzitskaya and Finkelstein
(GF)~\cite{Finkelstein}. In the GF model, in fact, each residue of the
polypeptide chain can be either in an ordered (native) or disordered
(non-native) state, encoded by the two possible values of a binary 
variable. This approach, similarly to ours, significantly narrows the 
conformational space, that consists of $2^N$ conformations only, for a
polymer with $N$ residues. This approach, while drastically reducing
the computation time, is in agreement with the Zimm-Bragg
model~\cite{Zimm}, widely employed to describe the helix-coil
transition in heteropolymers.  

 The outline of the paper
is as follows. In Section~\ref{model} we describe the  model and the
simulation procefdure; in Section~\ref{single} we study the equilbrium
behavio r of a single domain;  in
Section~\ref{fluctuations} we investigate the role of fluctuations in
the  reciprocal influence between successive unfolding events; in
Section~\ref{pulling-rate} we investigate the dependence of the
unfolding force on the pulling rate and we sketch a simple analytic
treatment of the unfolding probability in the limit of an extremely
high pulling rate; finally, in  Section~\ref{conclusions} we draw the
conclusions of our work. 

\section{The model} 
\label{model}

\begin{figure}[t]
\begin{center}
\includegraphics[width=\columnwidth,scale=1.5]{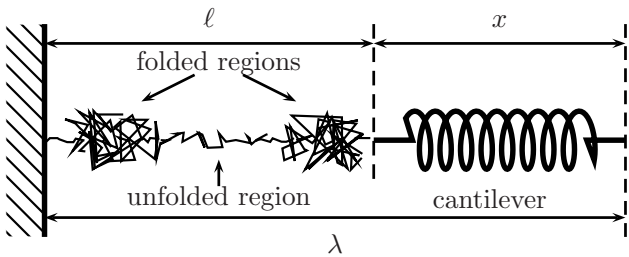}
\end{center}
\caption[Schematic description of the AFM experimental set-up.]{
Schematic description of the AFM experimental set-up. The polymer is composed 
by several tandem repeats of the same domain in series with a harmonic spring 
(the cantilever).
}
\label{setup}
\end{figure}

The AFM in its simplest arrangement is just a spring (the cantilever)
whose deflection, and thus the applied force, is measured as a 
function of its position. The system, as shown in Figure~\ref{setup},
can therefore be modeled as a harmonic spring in series with a 
protein composed by $M$ tandem repeats of the same domain. Each 
domain $j$ is simply portrayed as a sequence of contacts
($\boldsymbol{s}^{(j)}=\{s_i^{(j)}\}$, 
$i=1,\cdots,N$)  that can be  either intact ($s_i^{(j)} = 0$) or broken
($s_i^{(j)} = 1$), where $N$ is the total number of native contacts
inside each domain. The
length of the polymer chain can be simply computed as  
$\ell  
= \sum_{j=1}^M \sum_{i=1}^N s_{i}^{(j)}a =\sum_{j=1}^M a n_j $, 
 where $a$ represents the  incremental elongation
associated to each contact breakdown and $n_j = \sum_{i=1}^N
s_{i}^{(j)}$ is the number of broken contacts in domain $j$. 
 For the sake of simplicity, in
all our computations we set $a = 1$. The length of the spring, on the
other hand, is just $x = \lambda - \ell$ where $\lambda$ is
the extension of the  spring-polymer system, using the rest position
as the reference point.

The energy of a domain configuration $\boldsymbol{s}^{(j)}$ 
is modeled by the sum of two 
contributions: a harmonic term $H$,
 accounting for the presence of the spring,
and the sum over all  domains $j$ of a term $L^{(j)}$, 
related to the long-range interactions among the monomers, 
\[
 E = H + \sum_{j=1}^M L^{(j)}.
\]

The harmonic term is expressed as
\[ 
  H(x) = \frac{1}{2}Kx^{2} = \frac{1}{2}K(\lambda - \ell)^{2} = 
  \frac{1}{2}K\left(\lambda - a\sum_{j=1}^Mn_j\right)^{2},
\]
where $K$ is the harmonic constant. 

In our simplified  representation, if the fraction of intact contacts
$n_j/N$ in a 
domain $j$ is below a given threshold $\theta$, the domain receives a 
folding prize $AN$ proportional to the  number of possible contacts. 
The interaction term is thus computed as 
\[ 
  L^{(j)} = -AN\Theta(\theta N - n_{j}), 
\]
where $\Theta(x)$ is the Heaviside  step function
\[ 
  \Theta(x) = \left \{ \begin{array}{ll}
                         0 & \mbox{if $x < 0$} \\
                         1 & \mbox{otherwise}
                        \end{array} \right . 
\]

In summary, the energy of a configuration is just a function of 
the number of broken contacts in each domain 
\[
  E(\boldsymbol{n}) =  \frac{1}{2}K[x(\boldsymbol{n})]^{2}
   -AN\Theta(\theta N - n_{j}), 
\]
where $\boldsymbol{n} = \{n_1,n_2,\ldots,n_M\}$ and 
\[
  x(\boldsymbol{n})=\left(\lambda - a\sum_{j=1}^Mn_j\right).
\] 
This assumption speeds
up the computations, that may be performed in terms of the $n_j$. 

A stretching simulation starts from a completely folded initial
structure, where no contact is broken. The protein-spring length
$\lambda$, chosen as the control parameter, is linearly
increased from $\lambda_{Min}$ to $\lambda_{Max}$  in $k_{\lambda}+1$  steps
during the simulation;
\[
   \lambda(k) = \lambda_{Min} + \frac{\lambda_{Max} -
    \lambda_{Min}}{k_{\lambda}} k,
\]
where $k=0,1,\cdots,k_{\lambda}$.
For each value of $\lambda$, we compute the average length of the spring
$\langle x \rangle$ in an equilibrium simulation as
\[ 
  \langle x \rangle =
  \frac{1}{Z}\sum_{\boldsymbol{n}} 
  g(\boldsymbol{n})\; x(\boldsymbol{n}) e^{-\beta E(\boldsymbol{n})}, 
\]
where $Z$ is the partition function 
\[
  Z = \sum_{\boldsymbol{n}}g(\boldsymbol{n}) e^{-\beta
  E(\boldsymbol{n})},
\]
and the multeplicity factor
\[
g(\boldsymbol{n}) =\prod_{j=1}^M \binom{N}{n_j}
\]
is given in terms of 
the number of possible microscopic configurations containing 
$n_j$ cleaved contacts.   

In real stretching experiments, however, the polymer is subjected to a
finite pulling velocity, so that the molecule cannot be considered in
an equilibrium  condition. In order to consider this effect, we use
 Monte Carlo simulations. For each value of $\lambda$, $T$
Monte Carlo steps are performed, each involving $N\times M$ elementary
steps. The elementary step consists in the random choice of a  domain
and in the attempt to increase or decrease by one the number of
contacts with probability equal to the fraction of broken or intact
contacts respectively.  The trial move is then accepted or rejected
with a probability derived from the heat-bath criterion,
\[ 
  p(n_j \rightarrow n_j\pm 1) = \frac{e^{-\beta E(n_1,\dots,n_j\pm
  1,\dots,n_M)}}
{e^{-\beta E(n_1,\dots,n_j+1,\dots,n_M)} + e^{-\beta
E(n_1,\dots,n_j-1,\dots,n_M)}}, 
\]
where $\beta$ is the inverse temperature.

The average length  of the spring  $\langle x \rangle$, corresponding
to the current position  $\lambda$ is computed averaging over $T$
Monte Carlo steps.

\section{Results}

We first analyze the entropic effects related to temperature
through the analysis of computations in the absence of a folding
prize, and then investigate the role of long-range interaction by
setting a non-zero prize on a single-domain polymer. After that, 
we discuss a set of computations on a three-domain protein,
showing the importance of the coupling due to the cantilever 
in the mechanism of mechanical unfolding 
and, in particular, they explain how the first unfolding event affects the 
following ones. In the last part, we discuss the relation between pulling rate
and unfolding force, finding a logarithmic law. The section is completed with a
analytic treatment of the unfolding probability valid in the limit of high 
pulling rate.

\begin{table}
\begin{tabular}{|l|l|} \hline
$N$ : &  Maximum number of contacts per domain \\ \hline
$M$ : &  Number of domains \\ \hline
$L$ : &  Maximal length of the polymer-spring system \\ \hline
$k_{\lambda}$ : & Number of steps in the length of the polymer-spring system \\ \hline
$T$ : &  Number of Monte Carlo steps \\ \hline
$A$ : &  Folding prize (in units $N$) \\ \hline
$\theta$ : & Threshold to keep the folding prize \\ \hline
$K$ : &  Elastic constant of the cantilever spring \\ \hline
$\beta$ : &  Inverse temperature \\ \hline
\end{tabular}
\caption{Legend of the symbols appearing
in the figures.}
\label{legenda}
\end{table} 

The legend of the symbols appearing
in the figures is shown in Table~\ref{legenda}.

\subsection{Single domain analysis}
\label{single}

\begin{figure}[t]
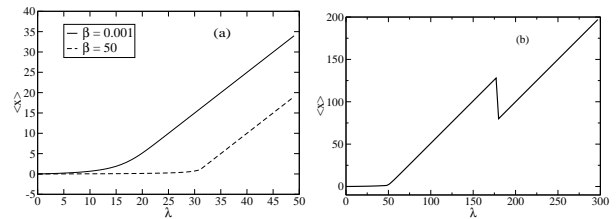
 
\begin{center} 
\begin{tabular}{cc}
\includegraphics[width=0.45\columnwidth]{no_prize_3P10_EQ.eps}&
\includegraphics[width=0.45\columnwidth]{prize_typical_plot_EQUIL.eps}
\end{tabular}
\end{center}
\caption[Cantilever deflection vs. estension for a single domain.]{
Cantilever deflection (force) versus estension for a single domain.
(a) No folding price. Influence of temperature-dependent entropic
effects on mechanical unfolding. Simulation
parameters: $N = 30$; $M = 1$; $L = 50$; $k_{\lambda} = 50$; 
$A = 0$; $K = 0.1$; $\beta = 0.001$ (solid line); $\beta = 50$ (dashed
line). (b) Typical ramp-like profile with folding prize. 
Simulation parameters: $N = 100$; $L = 300$; 
$k_{\lambda} = 100$; $A = 5$; $\theta = 0.5$; $K = 0.1$; $\beta = 2$.
}
\label{singlemodule}
\end{figure}

The force-extension
curves without folding prize, as shown in
Figure~\ref{singlemodule} (a), 
are clearly bi-phasic. The flat part of the
curve at low temperature ($\beta = 50$)
represents the complete unfolding of the  protein while the
spring nearly retains its resting length: 
at low temperatures, the enthalpic contribution of the free
energy (the harmonic energy of the cantilever), dominates over the
entropic one. 
When the
protein is completely stretched, the system can react to the increase
of the control parameter $\lambda$, only through an equal increase of
the spring length. The steep part of the $(x,\lambda)$ plot is thus a
straight line with unitary slope.

At high temperature ($\beta = 0.001$) the free energy is dominated by
the entropic term so that for small values  of $\lambda$, about $50
\percent$ of the monomers are extended in the direction of the pulling
force so as to maximize entropy, while the spring remains in  its
resting position. The proportion of unfolded monomers remains
thereafter almost unchanged during the simulation and for $\lambda >
15$, any further increase  in $\lambda$ is reflected in a equal
extension of the spring  $\Delta x = \Delta \lambda$.

\subsection{Effect of folding prize} 
\label{typical-plot}

The protein is here composed by a single domain
with $N = 100$ contacts, and its energy is lowered by $A = 5N$
when a fraction of residues greater than $\theta = 0.5$ is 
folded. 

The $(x,\lambda)$ plot portrayed in Figure~\ref{singlemodule} (b) shows a flat
region for  $\lambda \le 50$. This reflects the cleavage of $N \times
\theta = 50$ contacts that can occur without the loss of the folding
prize, while the spring remains very close to the resting length. As
the further extension of the protein would result in a significant
destabilization of the system due to the loss of the folding prize,
the increase of the control parameter $\lambda$ is now completely
accounted for by the stretching of the spring. The second part of the 
$(x,\lambda)$ plot is thus a straight line with unit slope. When the
increase in  harmonic energy exceeds the folding prize, the stretching
of the spring is interrupted and the remaining 50 contacts of the
protein break down, allowing a  corresponding shortening of the
spring. As the protein is now completely extended, any further
increase in $\lambda$ must result in a corresponding stretching of the
cantilever and the final part of the $(x,\lambda)$ plot is again a
straight line with unit slope.

\begin{figure}[t]
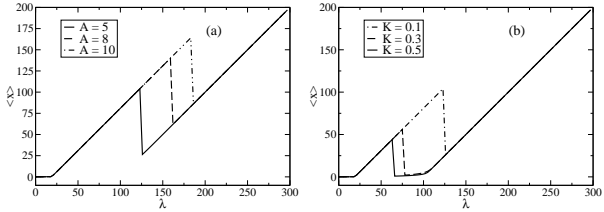

\begin{center}
\begin{tabular}{lll} 
\hspace{-1.5 cm} &
\includegraphics[width=0.45\columnwidth]{prize_A_005_008_01_EQUIL.eps} &
\includegraphics[width=0.45\columnwidth]{prize_K_01_03_05_EQUIL.eps}
\end{tabular}
\caption[Role of folding prize.]{
Role of folding prize $A$ (a) and harmonic constant $K$ (b). 
Common computation parameters: $N = 100$; $M = 1$; $L = 300$; 
$k_{\lambda} = 100$; $\theta = 0.20$; $\beta = 2$.
}
\label{AK}
\end{center}
\end{figure}

The features of the saw-tooth $(x,\lambda)$ profile are affected by
several simulation parameters. The folding prize $A$ is related to the
enthalpic component of the free energy and stabilizes the folded
conformation of the protein. As a consequence, when $A$ is large, the
polymer tends to remain in the compact conformation so as to retain
the significant  folding prize and the increase in $\lambda$ leads to
a stretching of the cantilever spring. Only when $\langle x \rangle$
is very large it becomes enthalpically favourable for the polymer to
unfold because the decrease in harmonic energy due to the cantilever
relaxation more than compensates the loss of the folding prize. Thus,
as $A$ is increased, higher and higher values of $\langle x \rangle$
are required for the elastic energy to compensate the folding prize
and the peak of the $(x,\lambda)$ plot becomes accordingly higher and 
higher. The harmonic constant $K$ of the cantilever spring plays a
role basically opposite  to that of the folding prize. In fact, when
$K$ is large, smaller average extensions $\langle x \rangle$ are
required for the harmonic energy to balance the folding prize so that
larger $K$s result in a less pronounced peak in the $(x,\lambda)$
plot. The role of $A$ and $K$ is exemplified by the simulations
portrayed in Figure~\ref{AK}.

\begin{figure}[t]
\begin{center}
\includegraphics[width=\columnwidth]{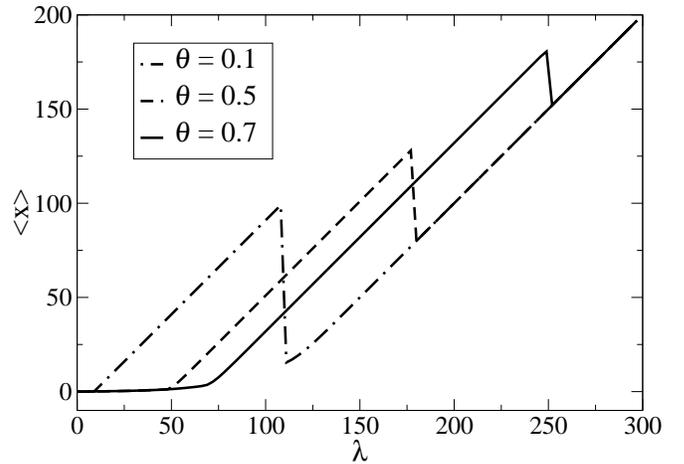}
\end{center}
\caption[Role of the threshold $\theta$.]{Role of the threshold $\theta$. Other computation parameters:  $N =
100$; $M = 1$;  $L = 300$; $k_{\lambda} = 100$; $A = 5$; $K =
0.1$; $\beta = 2$.
}
\label{theta}
\end{figure}

As just discussed, the role of the folding prize and of the harmonic
constant is related to the enthalpic term of the free energy. By
contrast, the folding threshold $\theta$ influences the entropic
contribution to the free energy. A small value of $\theta$ in fact,
implies that only a small fraction of the contacts can be broken
without loss of the folding prize. If $m < \theta N$ represents the
number of broken contacts  in a moment preceding the unfolding event, 
then the number of microscopic conformations allowed will be
$g(n,N)=\binom{N}{n}$,
and the entropy will be $S = -T\log[g(n,N)]$. If  $\theta < 0.5$, the
unraveling of the domain will  increase the degeneracy $g(n,N)$  and
thus the entropy, so that a moderate extension of the spring will be
sufficient for the enthalpic term to be more than compensated by the
entropic one.  By contrast, when $\theta \ge 0.5$, the breakdown of
the domain and the resulting increase in the number of disrupted
contacts, will bring the degeneracy and the  entropy further away from
the maximum thus disfavouring the unfolding event and  causing the
peak of the $(x,\lambda)$ profile to become higher. The pattern of
increase in height of the unfolding peak as $\theta$ takes on higher
values is shown in Figure~\ref{theta}. 

\begin{figure}[t]
\begin{center}
\includegraphics[width=\columnwidth]{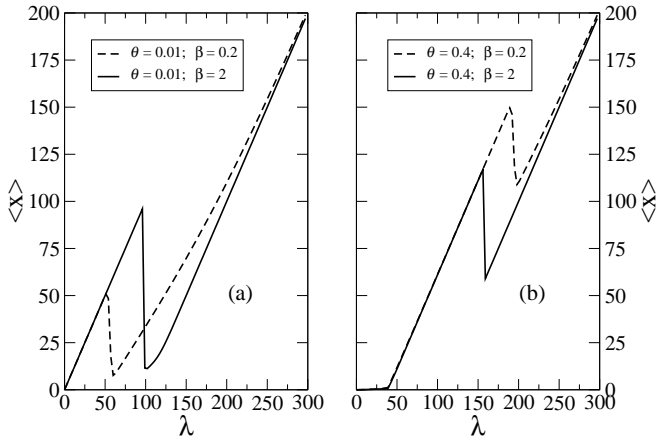}
\end{center}
\caption[Folding threshold and temperature.]{
Interplay between the folding threshold $\theta$ and the inverse
temperature $\beta$. For small  $\theta$s  (a) higher
temperatures favour the unfolding, whereas the opposite is true  for
large $\theta$s (b). Common computation parameters: $N =
100$; $M = 1$;   $L = 300$; $k_{\lambda} = 100$; $A = 5$; K =
0.1.
}
\label{thetabeta}
\end{figure}

The relevance of the entropic contribution on free energy computation
strongly  depends on temperature that may amplify the role of the
threshold $\theta$. As already noticed, in fact, when $\theta < 0.5$,
the breakdown of the domain is  entropically favoured as it brings the
degeneracy closer to its maximum.  Since this gain in entropy becomes
larger and larger as the temperature is increased, for small $\theta$
the unfolding event becomes more and more favourable as $\beta$ is
decreased, and the height of the peak of the $(x,\lambda)$ plot will
also decrease accordingly. For $\theta \ge 0.5$ the opposite pattern
can be observed. In fact, the entropy loss due to the decrease of the
degeneracy resulting from the unfolding event, is magnified as $\beta$
is decreased leading to higher and higher peaks in the $(x,\lambda)$
plot. The interplay between the parameters $\theta$ and $\beta$ is
shown in Figure~\ref{thetabeta}.

\subsection{Coupling and fluctuations}
\label{fluctuations}

\begin{figure}[t]
\hspace{-1 cm}
\includegraphics[width=\columnwidth]{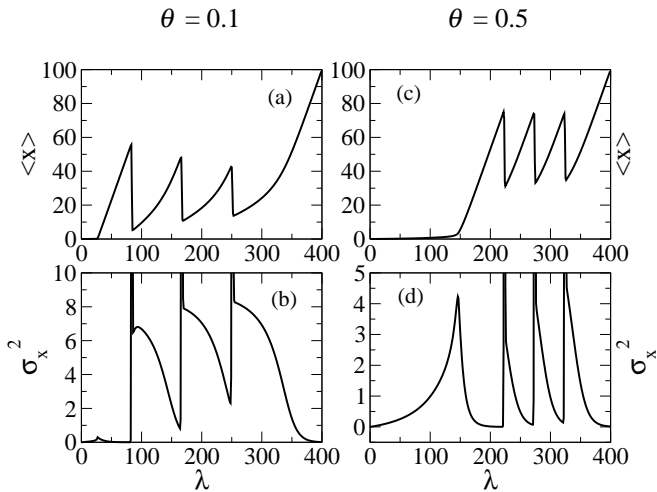}
\caption[Role of fluctuations.]{
Role of fluctuations. Panels (a) and (b): a threshold lower than 0.5
($\theta = 0.1$) causes the first  unfolding event to favour the
following ones. Panels (c) and (d): the first unfolding event does
not  affect the second and third ones as a result of the high
threshold $\theta = 0.5$. Computation  parameters: $N = 100$; $M = 3$;
 $L = 400$; $k_{\lambda} = 400$; $A = 1$; $K = 0.05$;  $\beta
= 2$.
}
\label{fig:fluctuations}  
\end{figure}

\begin{figure}[t]
\hspace{-1 cm}
\begin{center}
\includegraphics[width=\columnwidth]{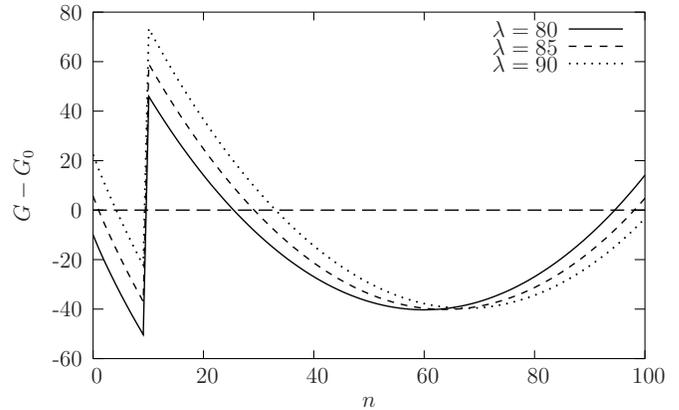}
\end{center}
\caption[Free energy landscape.]{
Free energy landscape under the action of an
increasing elastic force in  correspondence of the unfolding
peak shown in the top left panel of Figure~\ref{fig:fluctuations}.  
Simulation
parameters: $N = 100$; $M = 3$;  $L = 400$; $k_{\lambda} = 400$; $A =
1$;  $\theta = 0.1$; $K = 0.05$; $\beta = 2$.
}
\label{freeenergy}
\end{figure}

Before  studying the mutual influence of the
unfolding events, let us illustrate the features of the free energy
landscape of a single module near the unfolding transition. In
Figure~\ref{freeenergy} we show the evolution of the free
energy landscape  in correspondence of the unfolding transition
in a simulation with $M=3$ and $\theta = 0.1$  (see also top left panel of
Fig.~\ref{fig:fluctuations}). The profile is characterized by a cusp-like,
narrow well corresponding to the folded state, and a wide, smooth well
related to the unfolded state. The width of the two wells, in fact,
depends on the number of conformations that the system can explore: in
the folded state, the system is stretched due to the action of the
spring and no fluctuations are allowed so that only one conformation
will be populated; after the transition,  the residues of the
collapsed domain become free to fluctuate and many conformations  will
be explored thus determining a very wide well in the free energy
profile.  The Figure shows that for low values of $\lambda$ (and thus low
values of the elastic force),  the free energy of the  reference
folded state $G_0$ is lower than that of the unfolded  state $G$, thus
forbidding the breakdown of the domain; as $\lambda$ is increased, 
the free energy of the folded state increases, until it finally
becomes higher than  that of the folded conformation and the
stretching transition occurs. This is a typical example of a
first-order phase transition.

We now consider a polymer composed by $M = 3$ tandem repeats with $N =
100$ contacts each. In Figure~\ref{fig:fluctuations} we compare two simulations
performed with the same parameter set except for a different threshold
$\theta$. In the simulation with $\theta = 0.1$ the peaks
corresponding  to the second and third unfolding events are less
pronounced than the first one thus  suggesting that the unfolding of a
domain actually favours further unfolding events.  Conversely, in the
simulation with $\theta = 0.5$, the three peaks feature almost the
same height showing that the first unfolding event has little or no
influence at all on  the following ones.

Each unfolding event in Figure~\ref{fig:fluctuations} corresponds to a peak in
the variance  plot because the unraveling of a domain increases the
fluctuations of the polymer and spring length. The variance peaks  are
characterized by a high and narrow spike followed by a smoother region
that decreases more slowly. The shape of the variance peak is related
to the regions of the free energy landscape explored by the system
during the unfolding transition.  The spikes in the  variance plots
(that are truncated for the sake of graphical clarity) correspond to
the  situation with $G = G_{0}$ when both wells are explored by the
system and the variance  $\sigma_{x}^{2}$ is related to the distance
between the two  wells. On the other hand,  the smooth regions of the
variance plots refer to the case with $G < G_{0}$ when the system only
explores the unfolded region of the free  energy landscape whose width
correlates with the variance.

Figure~\ref{fig:fluctuations} shows that the height of the smooth region
of the variance peaks increases with the order of the unfolding event.
This trend is due to the fact that, with each unfolding event,
$N(1-\theta)$ new monomers become free to fluctuate and the number of
accessible  conformations increases accordingly.
Figure~\ref{fig:fluctuations} also  shows that the value of
the variance
$\sigma_{x}^{2}$ after each unfolding event, gradually decrease as
$\lambda$ is increased, because the disruption of the contacts of the 
domain just collapsed allow an extension of the molecule in the
direction of the  pulling force so as to avoid as far as possible  a
further stretching of the spring that would cause an increase in
energy. As a result, narrower and narrower regions of the conformation
space become accessible to the polymer and the variance is lowered. It
is worthwhile noticing, however, that the extension of the unfolded
domain is hindered by the subsequent decrease in entropy  so that the
number of contacts actually broken before each unfolding event is
smaller than the maximum number allowed by the loss of the folding
prize in the previous domain breakdown. 

When $\theta = 0.1$, 26 contacts (a number of the order of  $\theta
NM$) are broken before  the first unfolding event. This value is
consistent with the number of contacts that can be disrupted without
loss of the folding prize. After the first collapse event, the number
of contacts broken in the simulation rises to 115, thus determining an
increase of the fluctuations and favouring the breakdown of another
domain. This explains why the second peak of the $(x,\lambda)$ plot is
less pronounced than the first one. The occurrence of the second
unfolding determines a further increase of the number of cleaved
contacts to 205. This results in a  easy breakdown of the last domain
of the protein and the last peak of the $(x,\lambda)$ plot  is
therefore less high than the second one.

This scenario is significantly different for $\theta = 0.5$. For large
values of $\theta$, in fact, only a small number of residues can be
recruited for fluctuations after each unfolding event. As a
consequence, the variance $\sigma_{x}^{2}$ rapidly goes to zero after
each unfolding event and the fluctuations are extinguished before the
force threshold for unfolding can be crossed. The following unfolding
events, similarly to the first one, will occur when  the increase in
harmonic energy balances the loss of the folding prize and therefore
the height of the unfolding $\langle x \rangle$-peaks will be roughly
the same.  

\begin{figure}[t]
\hspace{-1 cm}
\includegraphics[width=\columnwidth]{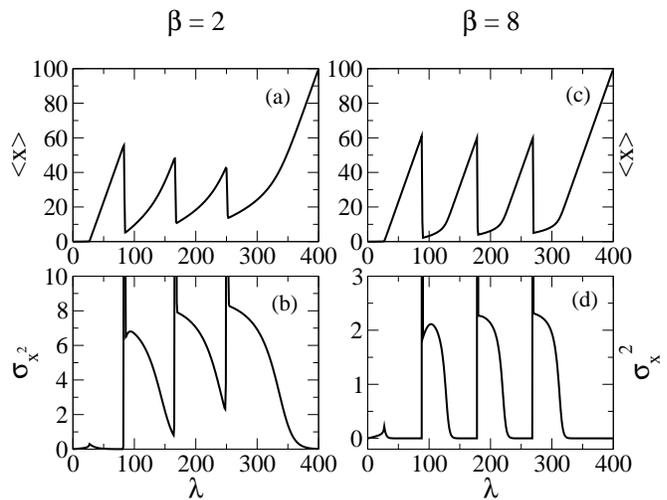}
\caption[Effect of temperature on the relative heights of the peaks.]{
Effect of temperature on the relative heights of the peaks of the
saw-tooth profile.  For $\theta = 0.1$, if the temperature is
sufficiently high ($\beta = 2$, panels (a) and (b)), the first
unfolding event favours the following ones due to the role of
fluctuations. At low temperatures however ($\beta = 8$, panels (c) and
(d) ), the fluctuations  rapidly become negligible and the heights of
the peaks become roughly the same due to the identical energetic
features of the domains. Computation parameters: $N = 100$; $M = 3$;
$L = 400$; $k_{\lambda} = 400$; $A = 1$; $\theta = 0.1$; $K =
0.05$.
}
\label{temperature}
\end{figure}

The role of fluctuations in determining the relative heights of the
peaks of the $(x,\lambda)$ plot, and thus the coupling among domains,
is confirmed through simulations performed at different temperatures.
At very low temperatures,  in fact, the entropic term of the free
energy becomes negligible and the height of the peaks of the
$(x,\lambda)$ plot only depends on the energetic term. As we are
considering a protein composed by identical domains, we expect the
$x(\lambda)$ peaks to be identical if the temperature is sufficiently
low. This pattern can be observed in Figure~\ref{temperature} where we
compare two simulations performed at different temperatures, namely
$\beta = 2$ and $\beta = 8$.

\subsection{Pulling rate effects}\label{pulling-rate}

\begin{figure}[t]
\begin{center}
\includegraphics[width=\columnwidth]{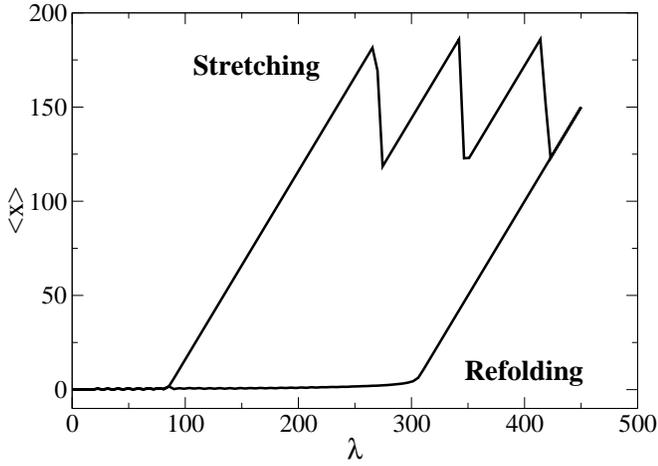}
\end{center}
\caption[Monte Carlo computations of stretching and refolding.]{
Monte Carlo computations of stretching and refolding: the
non-superposability of the  profiles is a signature of the
irreversibility of the process. Simulation parameters: $N = 100$; $M =
3$;  $L = 450$; $k_{\lambda} = 100$; $T = 1000$; $A = 1$; 
$\theta = 0.3$; $K = 0.5$; $\beta = 2$.
}
\label{montecarlo}
\end{figure}

Up to now, we discussed equilibrium stretching computations, 
\emph{i.e.} we assumed an infinitely slow pulling. However, at the
typical time scales of an AFM stretching experiment, the polymer is
pulled at a finite velocity and the unfolding is a non-equilibrium
process, as testified by the differences between the unfolding and the
refolding force-extension profiles (see e.g.\
Figure~\ref{montecarlo}).  In fact, while the unfolding profile features
the typical  saw-tooth pattern, upon relaxation of the unfolded
polymer, the trace exhibits no discontinuities that would indicate
refolding.

A consequence of the irreversibility of the stretching process is
that  the unfolding force depends on the pulling speed. Actually, if
the  loading rate $r_f = k_{f}v$ (with $k_{f}$ being the elastic
constant and $v$ the pulling velocity) is sufficiently small, thermal
fluctuations are allowed enough time to overcome the energy barrier
and the unfolding force will be low. 

Several experimental works~\cite{Ig, Carrion-veloc, Fritz-veloc,
Merkel-veloc}  reported a logarithmic dependence of the unfolding
force on the loading rate  in the case where a single energy barrier
is present along the reaction  path. The analytic expression of the
relation between force and loading rate was derived~\cite{Bell,
Evans-veloc, Tavan-veloc}  within the frame of Kramer's theory for a
simple two-state model, by
assuming that the external force reduces the  height of the energy
barrier,
\begin{equation} 
\langle F_{max} \rangle = \frac{k_{B}T}{\Delta x}
\log \left ( \frac{Kv \Delta x}{k_{0}k_{B}T} \right ),
\label{eq:evans} 
\end{equation}
where $k_B$ is Boltzmann constant, $T$ is the absolute
temperature, $\Delta x$ is the distance between the minimum
corresponding to the folded state and the activation barrier 
of the energy landscape, $v$ is the pulling speed, $K$ is the
cantilever harmonic constant and $k_{0}$ is the spontaneous
unfolding rate. 

Recently~\cite{Schlierf}, it has been shown that the probability
distribution of the force at various pulling rates does not follow the
simple Bell law, requiring the full Kramer's theory and predicting
small corrections to the logaritmic behavior of the most probable
force. 

\begin{figure}[t]
\begin{center}
\includegraphics[width=\columnwidth]{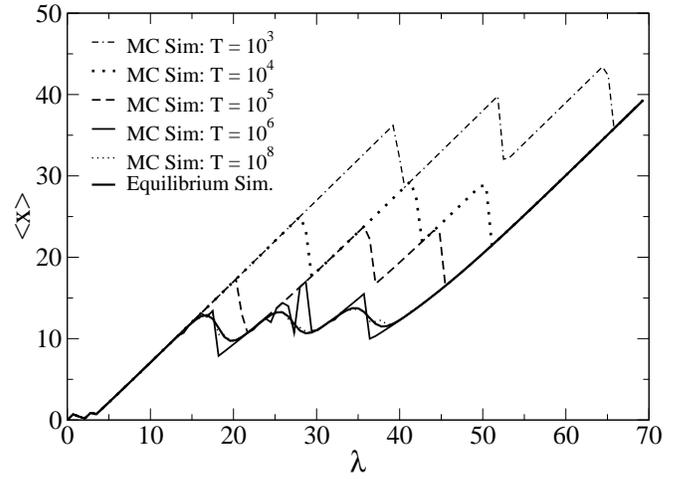}
\end{center}
\caption[Comparison between equilibrium and  Monte Carlo
simulations.]{
Comparison between the equilibrium computations and a set of Monte Carlo
simulations with  different numbers of MC steps. The peaks of the
$(x-\lambda)$ profile become lower and  lower and the MC simulations
converge to the equilibrium scenario as the number of MC  steps  is
increased. Simulation parameters: $N = 10$; $M = 3$; $L =
70$;  $k_{\lambda} = 100$; $A = 1$; $\theta = 0.2$; $K = 0.1$; $\beta
= 2$. Notice that, using a smaller value of $N$, the equilibrium
profile is smoother than that in  Figure~\ref{fig:fluctuations},
Figure~\ref{temperature} and preceding ones.
}
\label{dynamics}
\end{figure}

We limit here to a preliminary illustration of a
series of Monte Carlo simulations with
different number of steps $T$ considering the pulling speed as
being proportional to $1/T$.  A more detailed analysis will be
presented in a forthcoming paper.

Figure~\ref{dynamics} shows that, as the
number of Monte-Carlo (MC) steps is increased, the $(x-\lambda)$ plot
becomes closer and closer to the profile yielded by the equilibrium
simulation. In particular, a small number of MC steps, \emph{i.e.} a
fast pulling, corresponds to high peaks of the $(x-\lambda)$ profile,
whereas  larger numbers of steps correspond to less and less
pronounced peaks  and thus, smaller unfolding forces.

\begin{figure}[t]
\begin{center}
\includegraphics[height=\columnwidth,angle=-90]{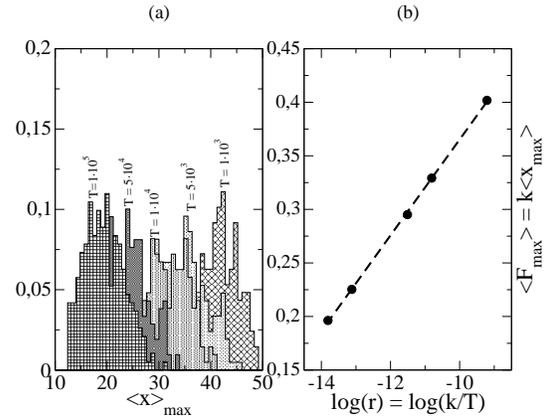}
\end{center}
\caption[Statistical analysis of unfolding forces.]{
Statistical analysis of unfolding forces in 5 sets of 100 independent
Monte Carlo runs  with different numbers of MC steps. Panel (a):
histograms of spring extensions at the rupture  point; panel (b):
linear fit of the average rupture force as a function of the logarithm
of  the loading rate. Simulation parameters: $N = 10$; $M = 3$; 
 $L = 70$;  $k_{\lambda} = 100$; $A = 10$;  $\theta = 0.2$; $K =
0.1$; $\beta = 2$.
}
\label{statistics}
\end{figure}

We performed a series of 100 independent
MC runs for 5 different values of $T$: $1\cdot10^{3}$,
$5\cdot10^{3}$,  $1\cdot10^{4}$, $5\cdot10^{4}$ and $1\cdot10^{5}$.
For each series of runs we built the histograms of the rupture spring
elongations and we plotted  the mean value of the histogram as a
function of the logarithm of the  loading rate. Finally, the set of
points thus obtained was linearly fitted.  As shown in
Figure~\ref{statistics}, the histograms, shift to lower values as the
number of MC steps increases. Figure~\ref{statistics} also shows that the 
mean rupture forces computed from the histograms, feature an
excellent  linear correlation with the logarithm of the loading rate
(correlation  coefficient $r_c = 0.99$).

If Eq.~\ref{eq:evans} is explicitly rearranged so as to show the
linear dependence of $\langle F_{max} \rangle$ on $\log(Kv)$, the
coefficients  of the linear equation can be equated to the
corresponding parameters $\gamma_{1}$ and $\gamma_{2}$ of the
regression line $\langle F_{max} \rangle = \gamma_{1}\log(Kv) +
\gamma_{2}$, so as to build the following system of equations:
\[ 
  \left\{ \begin{array}{lll}
            \frac{k_{B}T}{\Delta x} & = & \gamma_1 \\
            \frac{k_{B}T}{\Delta x}\log\left ( \frac{\Delta x}{k_{0}k_{B}T} \right ) & = & \gamma_2
            \end{array} \right . 
\]

From the first equation of the system it is possible to compute the
width  of the activation barrier $\Delta x = 11.17$; this value can
then be  substituted into the second equation so as to determine
$k_{0} = 2.93\cdot10^{-7}$.  The spontaneous unfolding rate $k_{0}$,
on turn, is related to the height of the activation barrier for the
unfolding process,
\[ 
  k_{0} = \omega e^{-\Delta G_{u}^{\dagger}/k_{B}T}, 
\]
where $\omega$, as explained by Kramer's theory, is the reciprocal of
a diffusive relaxation time. This simple computation shows how
stretching experiments and  simulations provide easy access to
important features of the free energy landscape.  In a forthcoming
paper we aim at investigating the relations relating the barrier width
$\Delta x$  and the spontaneous unfolding rate $k_{0}$ with molecular
properties such as the folding prize $A$, the threshold $\theta$, the
number $N$ and the length $M$ of the domains.

\subsection{High pulling rate limit}

Let us investigate the dependence of the
Monte Carlo simulations on the number of  Monte-Carlo steps $T$, 
that we can interpret as a measure of the pulling speed. 
In fact, in the limit of an
extremely high pulling speed, we can assume that for each value of
$\lambda$ the polymer  can adopt just a single (or at most, a few)
conformation, so that the entropic contribution can be neglected in
the computations.  
 
The mechanism outlined in Section~\ref{typical-plot} shows that the
key event  is the loss of the folding prize determining the complete
extension of the protein  domain and the subsequent relaxation of the
spring. The problem thus arises to  identify the factors affecting the
probability of this crucial step. More specifically,  suppose in a
domain a number of contacts just below the threshold to keep the
folding  prize has been broken. What is the probability $\pi$ that one
more contact will be  cleaved within the next $\nu$ steps ?

In order to answer this question, it is necessary to compute the
energy difference associated with the transition. Let $x$ be the
length of the spring when a number of contacts just below the
threshold $\theta N$ has been disrupted: the system retains the prize
$A$. If one more contact is broken, the prize is lost and the spring
will be correspondingly shortened. In particular, by assuming each
contact breakdown to cause a unit increase in length of the polymer,
the new length of the spring will  be $x-1$. The energy difference
between the two states can thus be computed as
\[ 
  \Delta E = \left ( \frac{1}{2}Kx^{2} - A \right ) - \frac{1}{2}K(x-1)^{2} 
  \cong Kx - A,
\]
where a term $K/2$ has been neglected. We can now compute the
probability to destroy a contact in a single step
\[ 
  p = \frac{1}{1 + e^{-\beta \Delta E}} = \frac{1}{1 + e^{\beta (A =
  Kx)}}.
\]

Finally, the probability to break a contact in $\nu$ steps is given by
the sum of a geometric progression
\[ 
  \pi = \sum_{i=0}^{\nu}(1 = p)^{i}p = 1 - (1 - p)^{\nu}. 
\]

The computations thus show that the domain unfolds with a probability
depending on the prize, the cantilever stiffness, the temperature and
the pulling velocity  (that in our simulations is related to the
number $T$ of Monte Carlo steps).

\begin{figure}[t]
\vspace{0.8 cm}
\begin{center}
\includegraphics[width=\columnwidth]{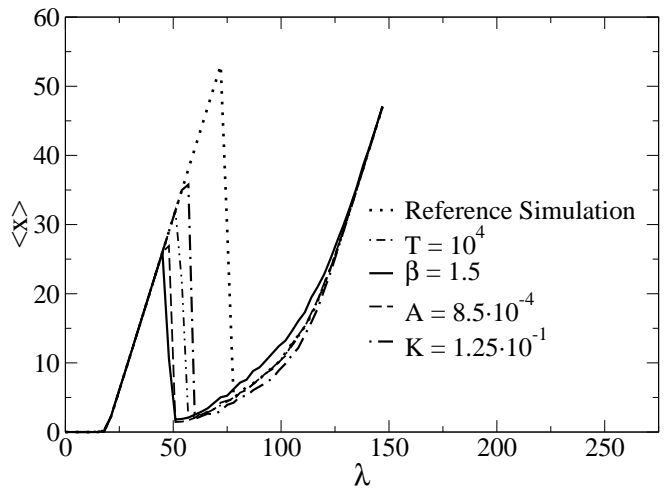}
\end{center}
\caption[Effect of the simulation parameters on the height of the
unfolding ramp.]{
Effect of the simulation parameters on the height of the unfolding ramp.
Unless otherwise indicated in the legend, the simulation parameters  are:
$N = 100$; $M = 1$; $L = 150$; $k_{\lambda} = 50$; $T = 100$;
$A = 0.1$; $\theta = 0.2$; $K = 0.1$; $\beta = 2$.
}
\label{height}
\end{figure}

In order to test our analytical treatment of the unfolding
probability, we performed a series of MC simulations differing from a
reference one just for a parameter  (Figure~\ref{height}). The key
parameters characterizing the reference simulations are: $T = 100$, $A
= 0.1$, $K = 0.1$, $\beta = 2$. In agreement with the analytic
computation, the simulations show that an increase in the number of
Monte Carlo steps ($T = 10^4$), an increase in temperature ($\beta =
1.5$), a decrease of the folding prize ($A = 8.5\times10^{-2}$), and
an increase of the spring constant ($K = 1.25\times10^{-1}$), all
result in an increase of the unfolding probability, thus reducing the
height of the peak in the     $(x-\lambda)$-plot.

\section{Conclusions}
\label{conclusions}

We developed an extremely simplified model of polymer stretching in
which the molecule is portrayed as a series of modules, represented as
an array of contacts,and a harmonic spring (the cantilever). The
chemical interactions stabilizing the  native conformation are simply
modeled as a folding prize gained by domains  where the fraction of
folded monomers is above a pre-chosen threshold. Our model is
consistent with recent findings in
Refs~\cite{Schulten1,Schulten2,Schulten3}, showing that the unraveling
of the titin Immunoglobulin domain occurs very rapidly only after the
breakdown  of a critical number of key hydrogen-bonds. The attribution
of a folding prize when a threshold value of the fraction of native
contacts is crossed, also makes our approach equivalent to a
G\={o}-model with force rescaling~\cite{Chan}.  However, our model is
thus significantly simpler than other models commonly used to study
mechanical unfolding such as the WLC and FJC models and the detailed
all-atom, topological and united-residue models employed for steered
molecular dynamics  simulations. Yet, our model is detailed enough to
reproduce many qualitative  features of the force-extension profiles
recorded in AFM experiments. In particular, our model correctly
reproduces the typical saw-tooth pattern with peaks characterized by a
height dependent on the folding prize, the temperature and the pulling
velocity.

In our study, a particular attention was paid to the relation between
the heights of successive peaks in the force-extension plot. This
point is quite intriguing as experimental data show that the force of
unfolding tends to  increase with each unfolding event, suggesting
that the protein domains unfold following an increasing order of
mechanical stability~\cite{Li2000}.  A possible explanation suggested
by the literature is that the domains of  the engineered modular
constructs used in AFM experiments are not identical but just
structurally  similar~\cite{Rief}. This argument may be correct in the
case of real proteins where the  differences in mechanical stability
of the tandem  repeats of the construct may arise from their different
position in the  tridimensional structure of the molecule. In a
computer simulation, however,  the domains are all perfectly identical
and the explanation for the staircase  pattern of unfolding peaks must
be sought elsewhere. This behavior has been ascribed to an independent
breaking probability of each domain. This hypothesis is however not
consistent with the fact that the cantilever couples the fluctuations
of all domains. Our unified treatment allows to keep into
consideration all contributions, and to obtain the correct equilibrium
curves, that generally do not exhibit this effect. However, a detailed
study of out-of-equilibrium extension curves, shows that this behavior
is consistent with a finite pulling speed, i.e.\ it can be ascribed to
a dynamical source.

Our results appear to be in agreement with recent findings by 
Cieplak~\cite{Cieplak} in steered molecular dynamics simulations of
titin  and calmodulin unfolding using a Go-like model. In
particular,it was found  that an increase in thermal fluctuations
results in a lowering of the force  peaks and in their earlier
occurrence during stretching.

An interesting feature of AFM stretching experiments is that the mean
unfolding force is a logarithmic function of the pulling rate. This
pattern reflects  the role of thermal fluctuations: if the pulling
speed is sufficiently low, then there will be enough time for thermal
fluctuations to drive the polymer over  the free energy barrier and
the unfolding force will be accordingly low. The  ability to reproduce
this pattern constitutes a benchmark for any stretching model. In
order to test our model, we performed a series of Monte Carlo
simulations with  different numbers $T$ of MC-steps and regarding the
pulling velocity to be proportional to $1/T$. Our toy-model, despite
its extreme simplicity, correctly reproduced the linear dependence of
the mean unfolding force on the logarithm of the pulling rate.

In summary, our simple model including harmonic and long-range energy
contributions, is not only capable of reproducing the force-extension
saw-tooth pattern, but it also yields force spectra displaying the
correct logarithmic dependence of  $\langle F_{max} \rangle$ on $Kv$
reported in experimental works. Finally, our model provides a simple
explanation of the influence of each unfolding event on the  following
ones, based on the role of fluctuations. Our work thus shows how
minimal models can be valuable tools even in the study of complex
molecular systems. 

\section*{acknowledgements}
We gratefully acknowledge fruitful conversations with Dr. L. Casetti
and Dr. M. Vassalli. We are also indebted with Dr. Vassalli for having
provided the experimental plot shown  in Figure~\ref{figVassalli}.
This work is part of the EC project EMBIO (EC contract n. 012835).


\end{document}